\documentclass[letterpaper]{jpconf}
\usepackage{graphicx}

\def\about{$\sim$}

\def\et{\emph{et al.}}
\def\apj{\emph{ApJ}}
\def\mnras{\emph{MNRAS}}

\begin{document}
\title{GLAST: physics goals and instrument status}

\author{Jennifer Carson}

\address{Stanford Linear Accelerator Center, MS 29, Menlo Park, CA 94025}

\ead{carson@slac.stanford.edu}

\begin{abstract}
The Gamma-ray Large Area Space Telescope (GLAST) is a space-based
observatory scheduled to launch in October 2007 with two instruments:
(1) the GLAST Burst Monitor (GBM), sensitive to photon energies
between 8 keV and 25 MeV and optimized to detect gamma-ray bursts, and
(2) the Large Area Telescope (LAT), sensitive to gamma rays between
~20 MeV and 300 GeV and designed to survey the gamma-ray sky with
unprecedented sensitivity. We describe the LAT and the GBM. We then
focus on the LAT's capabilities for studying active galactic nuclei.
\end{abstract}

\section{GLAST}

The Gamma-ray Large Area Space Telescope (GLAST) is scheduled to be
launched in October 2007 and will operate for 5-10 years in a
low-earth orbit. Unlike its predecessor, the Compton Gamma-ray
Observatory (CGRO), GLAST is not intended to make pointed
observations. Instead, it will operate primarily as an all-sky monitor
in which it continuously scans the sky, rocking $\pm 35^{\circ}$ about
the zenith every 90-minute orbit. GLAST will carry two instruments:

\begin{enumerate}

\item the Large Area Telescope (LAT), the main GLAST instrument,
sensitive to gamma rays between 20 MeV and 300 GeV, and

\item the GLAST Burst Monitor (GBM), dedicated to detecting gamma-ray
bursts (GRBs) between 8 keV and 25 MeV.

\end{enumerate}

\noindent Both instruments have completed all environmental testing and are
currently being integrated onto the spacecraft.

The GBM \cite{vonKienlin04} consists of twelve NaI crystal detectors
with sensitivity from 8 keV to 1 MeV and two BGO crystal detectors
with sensitivity from 150 keV to 30 MeV. The instrument has a field of
view of 9.5 sr (the entire sky not occulted by the Earth) and \about
$12\%$ energy resolution at 511 keV. The GBM is capable of on-board
localizations of $<15^{\circ}$ in 1.8 seconds and $2-3^{\circ}$ within
several seconds to a few minutes. It is anticipated to detect \about
200 GRBs per year, $> 50$ of which will be in the field of view of the
LAT.

The LAT \cite{michelson06} is a pair-conversion instrument. In each of
16 precision trackers, 14 layers of tungsten foil facilitate pair
conversion and 18 layers of X-Y pairs of single-sided silicon strip
detectors measure the pair tracks. The pair-initiated shower deposits
its energy in a calorimeter, composed of 1536 CsI crystals located at
the bottom of the LAT. A segmented array of plastic scintillators
surrounding the instrument detects charged particles as they enter and
is used to veto background events depending on energy and on the
correspondence of the hit scintillator tiles with tracks found in the
tracker.

Table \ref{LAT capabilities} \cite{lat_url} summarizes the LAT
performance. With a field of view of 2.4 sr, the LAT will ``see''
$20\%$ of the sky at any instant and will scan the entire sky once
every two orbits, or three hours. The predicted one-year sensitivity
is $F (E>100 \textrm{MeV}) > 3 \times 10^{-9} \textrm{cm}^{-2}
\textrm{s}^{-1}$ for a point source with a differential photon
spectrum proportional to $E^{-2}$ observed at high latitude. The
brightest point sources will be localized to $\sim0.4'$ and the
weakest sources to several arcminutes. The LAT will be much more
sensitive than its predecessor, the EGRET instrument aboard CGRO; in
one day, it will detect (at $5\sigma$) the weakest sources that EGRET
detected during the entire CGRO mission. The LAT is projected to
detect thousands of gamma-ray sources over the lifetime of the GLAST
mission.

\begin{center}
\begin{table}[h]
\caption{\label{LAT capabilities}LAT capabilities.}
\centering
\begin{tabular}{@{}*{7}{l}}
\br
Parameter&Present Design Value\\
\mr
Peak Effective Area&$10,000 \textrm{cm}^2$ at 10 GeV\\
Energy Resolution, 100 MeV, on-axis&$9\%$\\
Energy Resolution, 10-300 GeV, on-axis&$<15\%$\\
PSF, $68\%$, on-axis, 10 GeV (100 MeV) & $0.09^{\circ} (3.4^{\circ}) $\\
Field of view & $2.4 \textrm{sr}$\\
Source Location Determination & $< 0.4'$\\
\br
\end{tabular}
\end{table}
\end{center}

\section{Blazar physics with the GLAST LAT}

The LAT is expected to advance the scientific understanding of all
types of gamma-ray emitting objects, including Solar System sources
like the Sun and Moon, Galactic sources like supernova remnants and
pulsars, and extragalactic sources such as active galaxies and
GRBs. It will map the structured diffuse emission from the Milky Way
and will detect, or perhaps resolve, the diffuse extragalactic
emission as well. The LAT may also detect gamma rays from dark matter
annihilation and will almost certainly find new catagories of
gamma-ray sources. Each of these topics is covered in
\cite{michelson06}. Here we have chosen to concentrate on one type of
gamma-ray emitter, blazars, a population with significant scientific
overlap with ground-based TeV telescopes. We explore the potential of
LAT observations for understanding the physics of AGN jets.

\subsection{\label{sec: variability}Monitoring variability}

\begin{figure}[h]
\includegraphics[width=20pc]{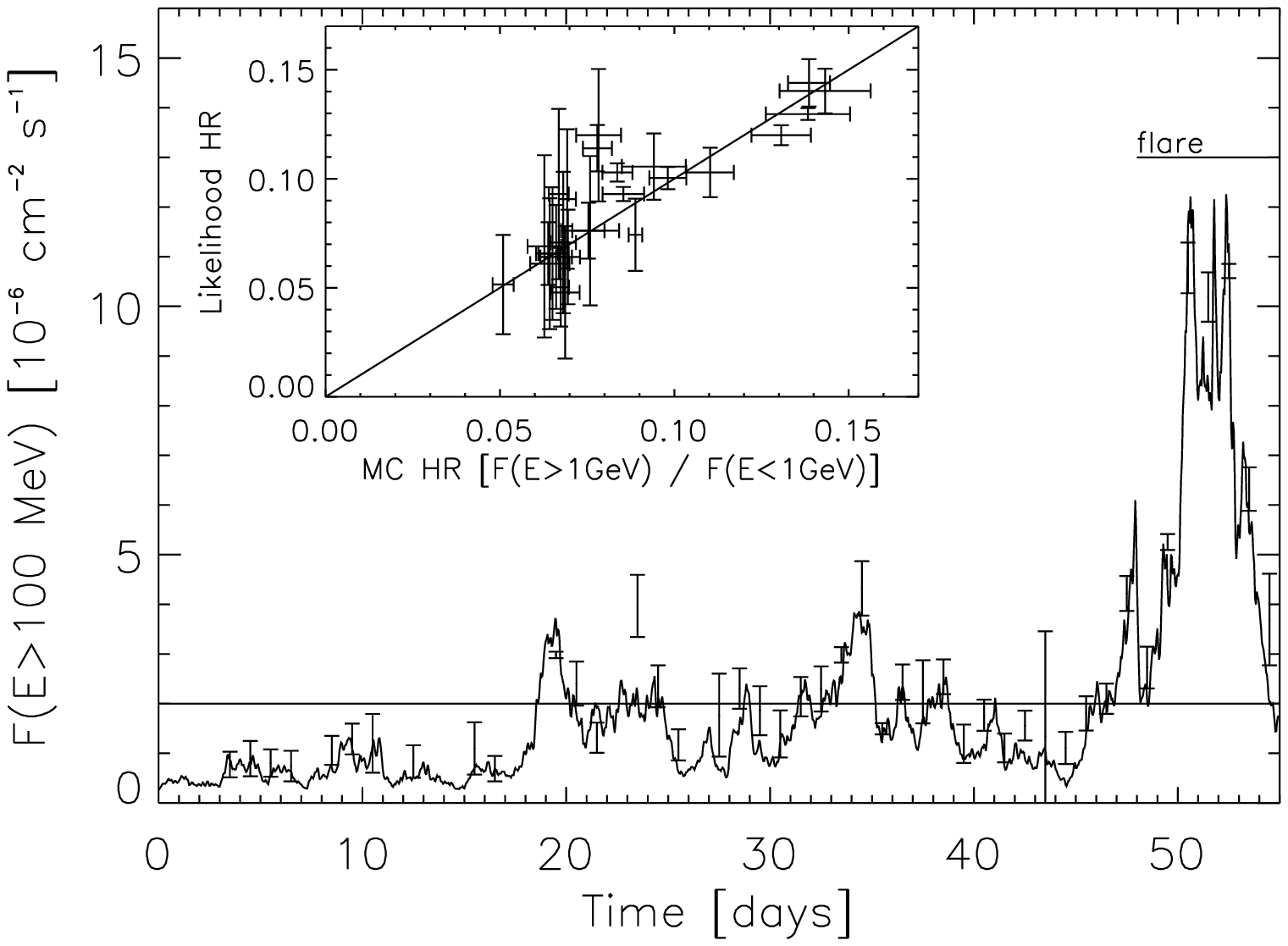}%
\includegraphics[width=20pc]{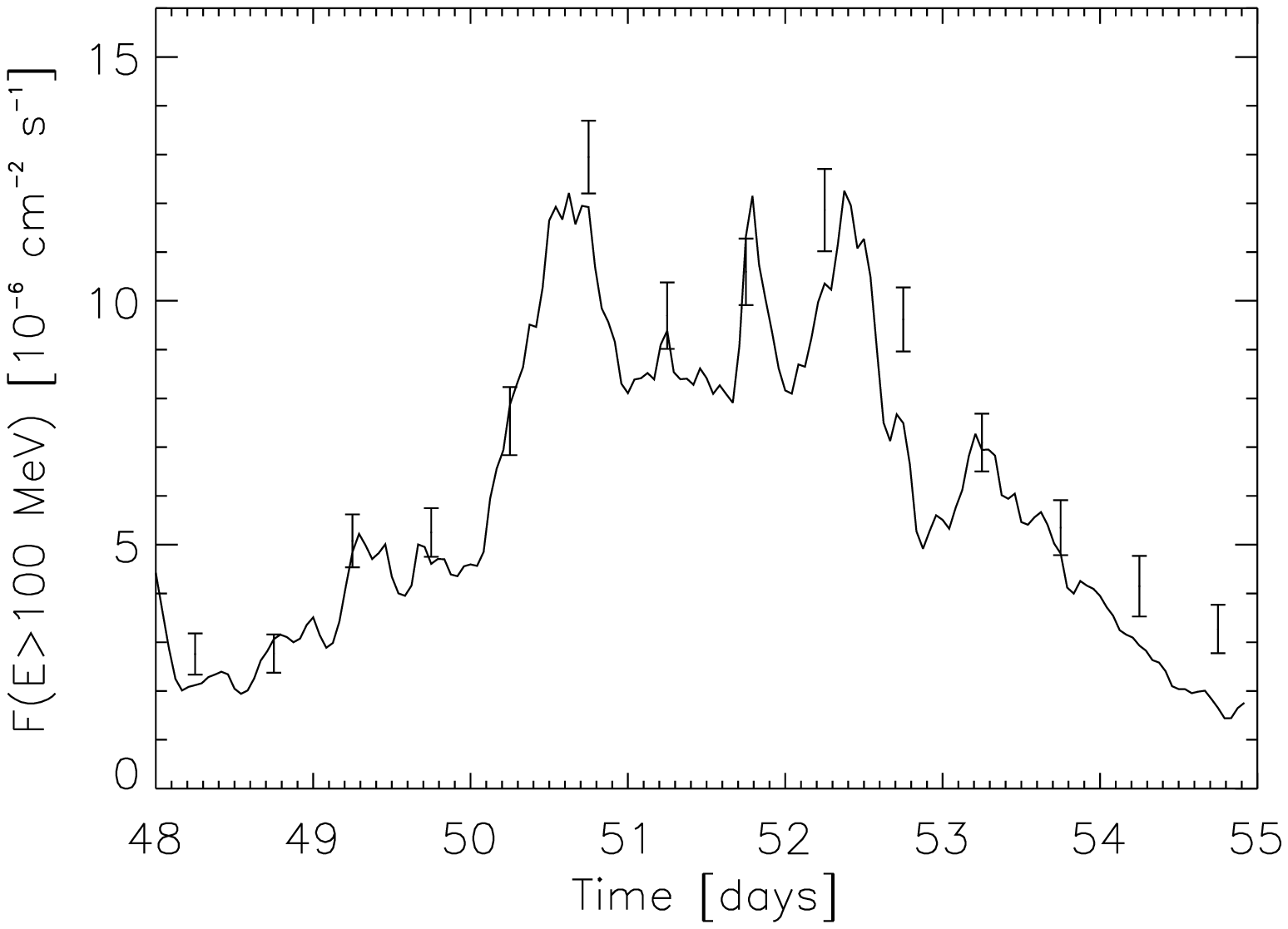}\hspace{2pc}%
\begin{minipage}[b]{40pc}\caption{\label{fig: variability}\emph{left}: A 55-day synthetic blazar light curve (solid line) plus one-day LAT exposures (data points). The inset shows the recovered vs. true hardness ratios. \emph{right}: Close-up of the flare indicated in the left panel, with LAT data in 12-hour exposures.}
\end{minipage}
\end{figure}

The frequency and uniformity of the sky coverage of the LAT will allow
sensitive, evenly-sampled monitoring of AGN variability across the
sky. Figure \ref{fig: variability} shows a 55-day synthetic light
curve that includes stochasic variability and a moderately bright
flare (solid line). The data points indicate fluxes derived for
one-day intervals from simulated LAT data. The data were analyzed
using an unbinned maximum likelihood technique that is being developed
as a standard analysis tool. The inset shows the hardness ratios ($F
(E ~ > 1 ~ \textrm{GeV}) / F (E ~ < 1 ~ \textrm{GeV})$) recovered from the
likelihood analysis vs. the true hardness ratios, indicating that
hardness ratios can be accurately measured on daily timescales, even
in low states. The horizontal line indicates the threshold for a
public data release in the first year; fluxes and flux ratios on any
object whose flux above 100 MeV exceeds $2 \times 10^{-6}
\textrm{cm}^{-2} \textrm{s}^{-1}$ will be released to the community
for follow-up observations and monitoring \cite{drp}. The right-hand
plot in Figure \ref{fig: variability} shows a close-up of the flare
with 12-hour time intervals. During moderate flares like the one
shown, fluxes can be measured to better than $10\%$ accuracy and
spectral indices can be measured to better than $5\%$ on 12-hour time
scales. Over the duration of the GLAST mission, the LAT is expected to
measure daily fluxes from thousands of sources with this level of
accuracy. Twelve-hour and hourly spectra can be measured for
approximately 100 and ten sources, respectively.

\subsection{Time-resolved spectral energy distributions}

The level of performance indicated in Section \ref{sec: variability}
suggests that the LAT will be able to measure the high-energy emission
from dozens of blazars on timescales of several hours. The synchrotron
cooling timescale for a population of relativistic electrons in the
inner jet can be several days for reasonable choices of the jet
parameters \cite{bottcher02}. Therefore, within the context of
leptonic models, 12-hour LAT spectra represent snapshots of the
particle distribution as it cools, and the LAT can track changes in
the gamma-ray spectral index as the highest-energy electrons
preferentially lose their energy to inverse-Compton scattering. In the
simplest SSC models, the free parameters are the magnetic field $B$,
the particle spectral index $p$ and upper- and lower- energy cutoff
$\gamma_1$ and $\gamma_2$, respectively, the size of the emitting
region $R$, and the bulk Lorentz factor $\Gamma$. Each LAT snapshot
constrains $B$, $p$, $\gamma_1$, and $\gamma_2$. If, in addition,
simultaneous X-ray observations that resolve the shortest variability
timescales are available, then these measure $R$ and $\Gamma$. The
X-ray spectral energy distributions (SEDs) also independently
constrain $B$, $p$, $\gamma_1$, and $\gamma_2$. We would expect the
constraint on $B$ to be particularly severe with such a set of
observations, if indeed it remains constant as the electron population
cools.

\subsection{Time-averaged SEDs}

\begin{figure}[h]
\includegraphics[width=20pc]{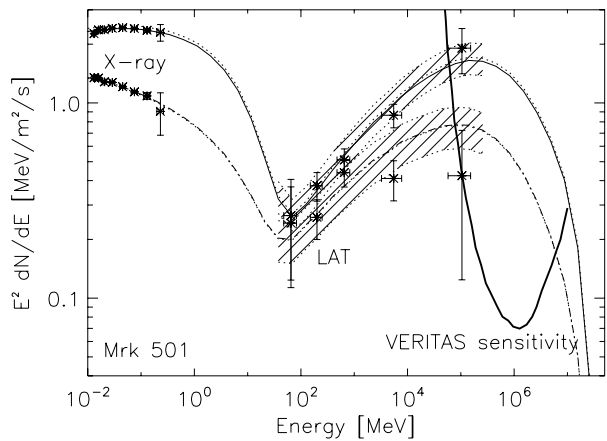}\hspace{2pc}%
\begin{minipage}[b]{20pc}\caption{\label{fig: mrk 501}The SSC models in high and medium states from \cite{petry00} (solid lines) are used to predict the LAT counts from a week of observations in survey mode. The points show the predicted LAT and X-ray counts from a binned likelihood analysis, and the shaded band indicates the $3\sigma$ LAT error from an unbinned likelihood analysis. The U-shaped line indicates the VERITAS sensitivity expected from 15 hours of observations (courtesy of R. Ong).}
\end{minipage}
\end{figure}

The estimated number of blazars that GLAST will detect ranges from at
least a thousand \cite{dermer06} to several thousand \cite{stecker96,
chiang98, mucke00}. The majority of these will be faint, and long
integration times will be required to build up a reasonable
high-energy SED. Here we explore the physics that can be probed with
SEDs that measure only the time-averaged properties of the jet. In
particular, we consider the case of a week of observations of
Markarian 501 (Mrk 501). In 1997, Mrk 501 was monitored by radio,
optical, X-ray ($2<E<12 ~ \textrm{keV,} ~ 20<E<200 ~ \textrm{keV}$),
and TeV ($E>800 ~ \textrm{GeV}$) telescopes simultaneously, and two
week-long epochs in medium and high states of activity were used to
fit SSC models \cite{petry00}. The modeling was realistic in that it
evolved the electron population self-consistently as it
cooled. Unfortunately, because no data existed on the rising edge of
the inverse-Compton peak, the models could not constrain $B$ or
$\gamma_1$, and so these parameters were fixed at nominal values. In
Figure \ref{fig: mrk 501}, we show the models for the 1997 medium- and
high-state epochs (solid lines). The X-ray points in Figure \ref{fig:
mrk 501} represent 25.2 ks (or 1 hour per day for a week) from a
BeppoSAX-like instrument; these cover the low-energy peak of the
SED. The gamma-ray points assume a week's worth of sky survey
observations with the LAT. 


As Figure \ref{fig: mrk 501} shows, joint LAT and and VERITAS
observations of Markarian 501, and of other high-frequency-peaked BL
Lac objects, will cover the entire high-energy peak of the SED. This
is an extremely powerful measurement for understanding the origin of
the high-energy emission, and such broad high-energy coverage will not
be possible until the launch of GLAST. In the context of leptonic
models, the LAT coverage of the low-energy half of the SED can
constrain $B$ and $\gamma_1$, unlike the previous modeling of
\cite{petry00}. If simultaneous X-ray data are also available that
cover the low-energy peak of the SED, then the overall energetics of
the inner jet are known. We can directly measure the relative
contributions of synchrotron and inverse-Compton cooling in the
jet. This type of complete, simultaneous coverage constrains all of
the parameters of simple SSC models: $B$, $p$, $\gamma_1$, $\gamma_2$,
$\Gamma$, and $R$.

\section{Conclusions}

We have described the two GLAST instruments and explored the
constraints that LAT observations can make on leptonic emission models
of AGN jets. We emphasize that none of the results shown require
pointed LAT observations; they are all achievable with the all-sky
scanning mode of observing. Of course, the most interesting findings
may be from sources where the LAT data rule out a simple SSC
picture. In these cases, either more complicated leptonic modeling or
hadronic modeling must be invoked. Finally, it is clear from the
examples here that in order to optimize the scientific return of GLAST
for blazars, simultaneous multi-wavelength data are essential,
especially from X-ray satellites and from TeV instruments such as
VERITAS and H.E.S.S.

\end{document}